\documentclass[conference]{IEEEtran}
\usepackage[dvips]{graphicx}
\ifCLASSINFOpdf
\else
\fi
\newcommand{\argmax}{\mathop{\rm argmax}\limits}
\usepackage[cmex10]{amsmath}

\begin{document}
%
\title{
Recoverable DTN Routing based on a Relay of\\
 Cyclic Message-Ferries on a MSQ Network
}

\author{\IEEEauthorblockN{Yukio Hayashi}
\IEEEauthorblockA{Graduate School of Knowledge Science\\
Japan Advanced Institute of Science and Technology\\
Nomi-city, Ishikawa-pref., Japan\\
Email: yhayashi@jaist.ac.jp}
}


%


\maketitle

\begin{abstract}
An interrelation between a topological design of network and 
efficient algorithm on it is important for its applications to 
communication or transportation systems.
In this paper, 
we propose a design principle for 
a reliable routing in a store-carry-forward manner 
based on autonomously moving message-ferries on a special structure 
of fractal-like network, which consists of a self-similar tiling of 
equilateral triangles.
As a collective adaptive mechanism, 
the routing is realized by a relay of cyclic message-ferries 
corresponded to a concatenation of the triangle cycles 
and using some good properties of the network structure.
It is recoverable for local accidents in the hierarchical 
network structure.
Moreover, 
the design principle is theoretically supported 
with a calculation method for the optimal service rates of 
message-ferries derived from a tandem queue model
for stochastic processes on a chain of edges in the network. 
These results obtained from a combination of 
complex network science and computer science will be useful for 
developing a resilient network system.
\end{abstract}


%
\IEEEpeerreviewmaketitle

\section{Introduction}
A delay/disruption-tolerant network (DTN) routing in a 
store-carry-forward manner is useful especially 
for disasters, battlefields, and poor communication 
environments on mobile devices or ad hoc wireless connections 
\cite{DSouza10,Shah11,Zhao05}.
Even if some connections are removed by failures or attacks, 
it is expected that a packet is delivered before very long 
in a DTN routing 
by re-establishing or re-generating an end-to-end route.
There are many protocols 
in DTN routing methods \cite{DSouza10,Shah11}. 
We focus on a message-ferries scheme \cite{Zhao05} 
instead of the most trivial approaches by flooding with 
a lot of redundant messages. 
In the message-ferries scheme, 
ferries of communication agents move proactively 
to send and receive messages, 
and distributedly give support to the delivery of messages. 
In the DTN as a collective adaptive system (CAS), 
it is very important how to design the interactions 
among message-ferries in what type of moves or actions.

We have proposed the searching and routing methods by 
message-ferries where actions obey random walks on a 
fractal-like network \cite{Hayashi13}.
The methods are better than the conventional optimal 
search by biophysically inspired 
L\'{e}vy flights on a square lattice
\cite{Viswanathan99,Santos05,Viswanathan11} 
for homogeneously distributed targets at unknown positions, 
because the fractal-like network structure is adaptive to 
the uneven distribution of 
communication flows between send and receive 
requests according to spatially 
inhomogeneous population densities.
However, random walks are predominated by 
contingency.  
For a higher reliable DTN routing, 
we consider a design principle of 
configuration and autonomous actions with recoverable procedures 
in CAS. 
It is based on the cooperation in cyclic moving of 
message-ferries on the multi-scale quartered (MSQ) network 
taking into account several advantages of the special 
fractal-like structure \cite{Hayashi09,Hayashi10}. 
Moreover, we give an estimation procedure for 
the optimal service rates of message-ferries 
from a queueing theory.
Thus, we focus on 
consistent simple ways for a collective (routing) function, 
theoretical and algorithmic validity, and adaptiveness for a change, 
which depend on the configuration and actions in CAS.

\section{Multi-Scale Quartered Network}
We consider a geographical network construction, in which 
the spatial distribution of nodes is naturally determined 
according to population in a self-organized manner. 
The following MSQ network model is generated 
by a self-similar tiling of faces
for load balancing of communication 
requests in the territories of nodes.
The territory is assigned by the nearest access from 
each user's position to a node on a geographical map.
Note that 
communication requests are usually more often generated and 
received at a node whose assigned population is large in the 
territory, 
and that the territories of nodes are determined by 
the nearest access.
Thus, how to locate nodes is important for 
balancing the communication load as even as possible.

\begin{figure}[htb]
\centering
  \includegraphics[height=67mm]{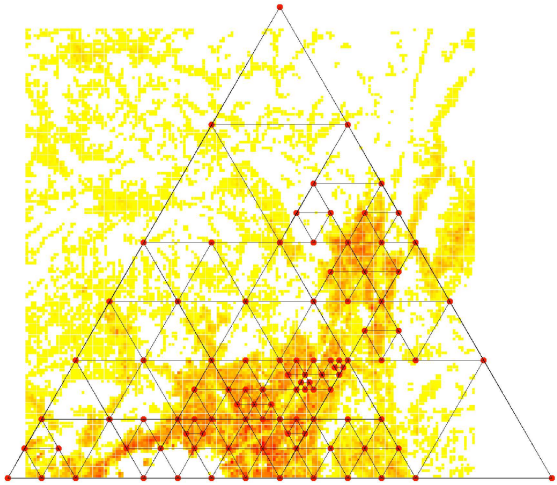}
\caption{Example of MSQ network generated by the subdivision 
of equilateral triangles on a geographical map. 
From white, yellow, to orange, 
the gradation is proportional to the population density.}
\label{fig_vis_msq}
\end{figure}

[Generation procedure of a MSQ network]
\begin{description}
  \item[Step0: ] Set an initial triangulation of any polygonal region 
    which consists of equilateral triangles.
  \item[Step1: ]  At each time $1, 2, 3, \ldots$, 
    a triangle face is chosen with a probability proportional to 
    the population in the space.
  \item[Step2: ] As shown in Fig. \ref{fig_vis_msq}, 
    four smaller equilateral triangles are created 
    from the subdivision by adding three nodes, 
    at the intermediate point of each edge of the chosen triangle, respectively.
  \item[Step3: ] Return to Step 1, while the network size 
	     (the total number of nodes) does not exceed a given size.
\end{description}

Since a step-by-step selection of triangle face 
is unnecessary in the above algorithm, 
the subdivision process can be initiated in a asynchronously 
distributed manner, e.g. according to 
the increase of communication requests in an individual triangle area.
We will discuss why the configuration of equilateral triangles 
is better in the next section. 
On a combination of complex network science and computer science 
approaches, this model has several advantages 
\cite{Hayashi09,Hayashi10}: 
the robustness of connectivity without vulnerable high degree nodes, 
the bounded short distance path between any two nodes, 
and the efficient decentralized routing on a planar graph 
which tends to be avoided from interference among wireless beams.
Complex network science that emerges at the beginning of the 21st century 
provides a new paradigm of network self-organization, 
e.g. based on recursive growing geometric 
rule for the division of a face 
\cite{Zhang08,Zhou05,Zhang06,Doye05}
or for the attachment which aims at a chosen edge
\cite{Wang06a,Rozenfeld07,Dorogovtsev02}
in a random or hierarchical selection.
These self-organized networks including the MSQ network 
have a potential to be superior to 
the vulnerable scale-free network structure 
found in many real network systems \cite{Albert02,Newman03}, 
such as Internet, WWW, power-grids, airline networks, 
etc.

\section{Reliable message-ferries routing}
\subsection{An efficient DTN routing}
\begin{figure}[htb]
\centering
  \includegraphics[height=62mm]{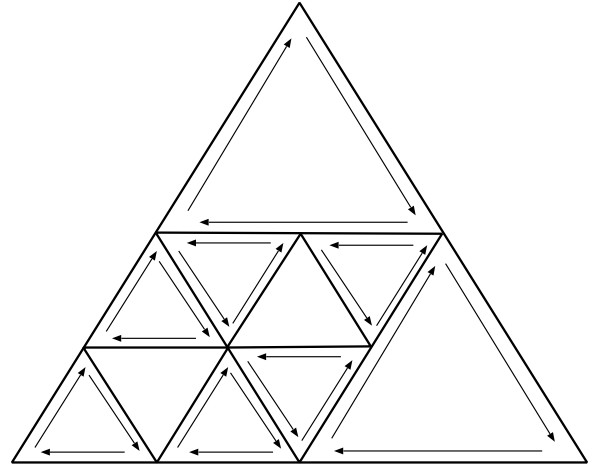}
\caption{Cycles on equilateral triangle faces 
in a MSQ network with the 1st and 2nd layers, 
which consist of large and small triangles, respectively.}
\label{fig_msq_cycle}
\end{figure}

\begin{figure}[htb]
\centering
  \includegraphics[height=62mm]{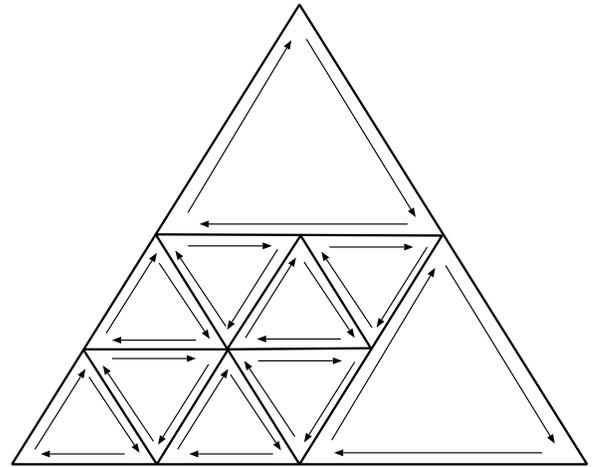}
\caption{Another idea of directional edges 
defined by only clockwise cycles.}
\label{fig_another_idea}
\end{figure}

We preliminarily explain some basic elements and mechanisms 
to realize a message-ferries routing on a planar network. 
Figure \ref{fig_msq_cycle} shows 
the forward direction of each edge defined by the clockwise cycles
on upper triangles and the counterclockwise cycles on lower triangles.
In order to 
reduce the number of assignment of a direction to each edge, 
we omit the assignment for the center faces of triangles 
at any layer. 
The layer is defined by a depth of quartered triangle faces 
(or by the number of the recursive divisions) from an initial 
outermost triangle. 
The backward direction of edge is defined in the same way 
by changing each of cycle to the opposite direction. 
We remark that 
a cycle is corresponded to three edges on the triangle face.
Moreover, 
these cycles represent the delivery routes of message-ferries.
We consider another idea as shown in Fig. \ref{fig_another_idea}: 
each triangle face is directed by only clockwise cycles.
Then, both forward and backward directions of an edge are assigned 
without counterclockwise cycles, except the edges of 
the outermost triangle.
The better case of either
Fig. \ref{fig_msq_cycle} or \ref{fig_another_idea}
depends on situations of the utilization in what type of communication 
resource and environment is given.

We consider the autonomous distributed delivery processes 
after a communication request is occurred at a source node.
In a store-carry-forward manner, 
a ferry that visits the source node $s$ picks up a data from $s$, 
and carries it to the next mediator node on the routing path, 
which is found by a routing algorithm as mentioned later.
The mediator node stores the data into its queue.
Here, the pickup and store times are ignored. 
Then, 
any other ferry that visits the 1st mediator node 
picks up the data, 
and carries it to the 2nd mediator node.
Such relaying by ferries are repeated until a ferry reaches at 
the terminal node $t$ and the data is delivered.
We assume that each ferry autonomously moves on a triangle cycle 
at a turnaround rate.
Several ferries exist on a cycle at random interval each other, 
and may have various speeds.
Therefore, in the random process by heterogeneous message-ferries, 
similar pickup and carry services are available at any time.
This property of the random process 
rationalizes an exponential distribution of 
service times in the next subsection. 
In this routing, 
direct interactions between ferries at a time
are not necessary, 
because a node act as the helper to temporarily store 
and forward a data asynchronously with ferry's encounters.
Note that our approach is categorized as 
multi-route and node relaying type 
in the ferry route design algorithms \cite{Zhao05}. 
Although the message ferries act just like information frames 
in Token Ring \cite{token_ring} at the date-link layer, 
it is quite different that 
the special topology of MSQ network generated by subdivision 
of equilateral triangles is focused therefore 
a relay of data transfers on cycles by mobile agents 
can be performed as a routing.

\begin{figure}[htb]
\centering
  \includegraphics[height=65mm]{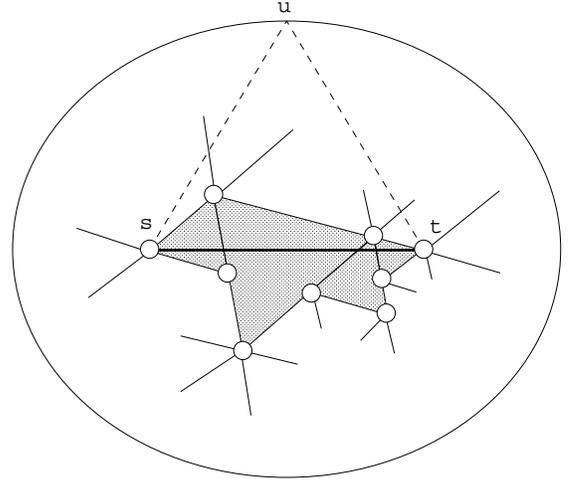}
\caption{Local search area in the ellipsoid with focus points 
$s$ and $t$ for the adaptive face routing. 
The chord length that consists of $s$-$u$ and $u$-$t$ is defined by 
the twice of $s$-$t$ line because of $t=2$-spanner.}
\label{fig_adaptive_face}
\end{figure}

\begin{figure}[htb]
\centering
  \includegraphics[height=65mm]{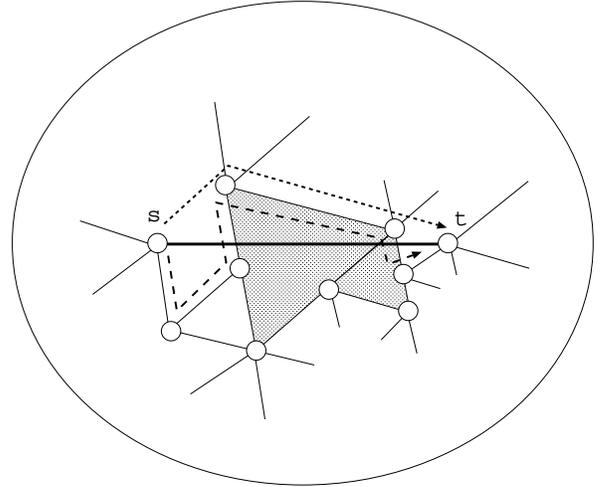}
\caption{Detour case in disconnections.}
\label{fig_detour}
\end{figure}

Since the MSQ networks 
that consist of squares and equilateral triangles 
belong to planner graps
with the $t=2$-spanner property \cite{Hayashi10,Karavelas01}, 
we can apply an efficient 
adaptive face routing algorithm \cite{Kuhn02,Bose04}
in order to find a shortest distance path between any two node.
The spanner property means that 
a length of path measured by 
the sum of link lengths on the path 
as Euclidean distances is bounded at most 
the twice of the straight line between two nodes. 
Efficiently, 
the routing algorithm uses only local information 
about the edges of faces that intersects the straight
line between source $s$ and terminal $t$ nodes, 
and the nodes of the faces are restricted in the ellipsoid 
whose chord length is defined by twice of the $s$-$t$ line 
as shown in Fig. \ref{fig_adaptive_face}. 
We remark that 
the routing path can be represented by 
a concatenation of triangle cycles for the intersected faces
with the $s$-$t$ line.
If there are several shortest distance paths with a same path length 
between two nodes, e.g. due to symmetry, 
each of the paths is equally selected at random with the rate 
1/(the number of the paths) $\times$ (the amount of communication 
flows between the nodes). 
Figure \ref{fig_detour} shows that the routing algorithm is able 
to find an alternative path when some edges are disconnected,
although a greedy forwarding policy based on the distance from the 
neighbor of a current node to the terminal node supports the path 
finding in 
the parts of removed edges and the corresponding triangle faces.
The dashed piecewise linear line 
denotes the shortest distance path in this case, 
and the dotted piecewise linear line 
denotes the original path.
Thus, even with several damages for the network, 
the routing can be performed 
in a connected part before the fully destruction.
In addition, 
the path finding is able to be performed reactively on-demand 
after a communication request occurs, 
and therefore adaptive for changes of connections.

\subsection{Optimal service rates of message-ferries}
We naturally assume 
that the occurrence of communication request 
is independent random event, 
but the amount is proportional to a product of populations around 
source and terminal nodes, 
and that the service to transfer the request is stochastic 
in a multihop manner by 
passing through mediator nodes on a network.
Thus, we apply a queueing model to the communication flows 
as follows.

For the M/M/1 queue system \cite{Chee-Hock08}    
as Kendall notation with the 
arrival rate $\lambda$ according to a Poisson process 
and the service rate $\mu$ in an exponential distribution, 
the average number of remaining requests is 
\[
  L = \frac{\rho}{1 - \rho}, \;\;\; 
  \rho \stackrel{\rm def}{=} \frac{\lambda}{\mu}, 
\]
and 
the average end-to-end delay is 
\begin{equation}
  E(T) = \frac{L}{\lambda} = \frac{1}{\mu - \lambda}.
\label{eq_ave_e-t-e_delay}
\end{equation}

\begin{figure}[htb]
\centering
  \includegraphics[height=15mm]{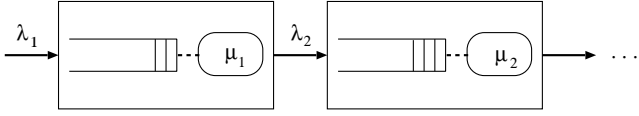}
\caption{Tandem queue model for a path.
The average end-to-end delay is given by 
$\sum_{i} \frac{1}{\mu_{i} - \lambda_{i}}$. 
}
\label{fig_queueing_model}
\end{figure}

\begin{figure}[htb]
\centering
  \includegraphics[height=55mm]{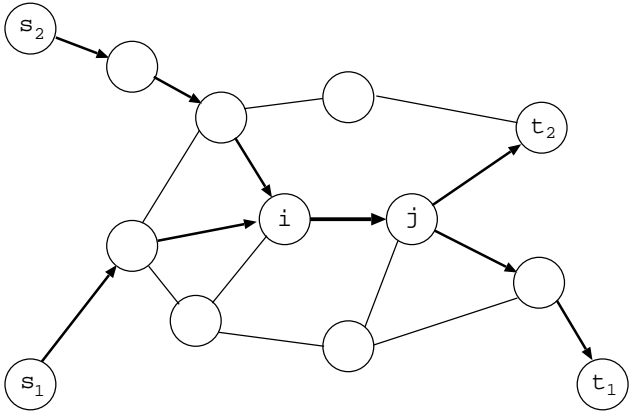}
\caption{Communication flows for an edge $i$-$j$ on two paths. 
Here, 
$s_{1}$ and $s_{2}$ denote source nodes, 
$t_{1}$ and $t_{2}$ denote terminal nodes.
The edge $i$-$j$ receives the directional 
flows superimposed by the amounts passing through the edge 
for paths $s_{1} \rightarrow t_{1}$, $s_{2} \rightarrow t_{2}$, 
and so on.
}
\label{fig_comm_flows}
\end{figure}

When we consider such stochastic processes on a network, 
the tandem queue model in Fig. \ref{fig_queueing_model} is applied. 
Each path between source and terminal nodes 
(e.g. in Fig. \ref{fig_comm_flows}) is decomposed into 
chained edges corresponded to M/M/1 queues. 
In the tandem queue model, from Burke's theorem, 
these queues are independent of each other in a Poisson process. 
Then, the average end-to-end delay in a direction 
is given by the sum of ether 
\[
\frac{1}{\mu_{l}^{f} - \lambda_{l}^{f}(k)}
\;\;\; {\rm or} \;\;\;
\frac{1}{\mu_{l}^{b} - \lambda_{l}^{b}(k)}
\]
in Eq.(\ref{eq_ave_e-t-e_delay}), 
where $\lambda_{l}^{f}(k)$ denotes the arrival rate for 
the sum of flows passing through an edge 
corresponded to the $l$-th cycle of message-ferry, 
the superscript $^{f}$ and $^{b}$
denote the forward and the backward directions of edge, 
the index number 
$k = 1, 2, 3$ is due to the one-to-three corresponding between 
a cycle and three edges of triangle.
If two cycles get involved in a same direction 
on an edge with respect to both adjacent sides of faces, 
the arrival rate $\lambda_{l}^{f}(k)$ per cycle is given by 
(the sum of flows)/$2$. 
We assume 
$\lambda_{l}^{f}(k) = \lambda_{l}^{b}(k)$ by the symmetry of 
flows to simplify the discussion.
The amount of flows on an edge is corresponded to the 
link load called as routing betweenness centrality \cite{Dolev10}. 
The service rate $\mu_{l}^{f}$ corresponds to the turnaround rate 
of the $l$-th cycle of 
message-ferry whose direction of clockwise or 
counterclockwise is determined by the coincidence 
with the forward direction of edge on the triangle cycle (see 
Fig. \ref{fig_msq_cycle}).
On each cycle, the condition 
\begin{equation}
  \mu_{l}^{f} > \max \{ \lambda_{l}^{f}(1), 
  \lambda_{l}^{f}(2), \lambda_{l}^{f}(3) \},
\label{eq_condition}
\end{equation}
is necessary for the stable situation 
without involving $\infty$-length queue.

We consider the delivery cost $C$ that consists of the end-to-end 
delay and the service load.
\begin{eqnarray}
  \sum_{i,j} \sum_{l \in path(i,j)} 
  \left( \frac{1}{\mu_{l}^{f} - \lambda_{l}^{f}(1)}
  \oplus \frac{1}{\mu_{l}^{f} - \lambda_{l}^{f}(2)}
  \oplus \frac{1}{\mu_{l}^{f} - \lambda_{l}^{f}(3)} \right. \label{eq_cost} \\ 
  \left.
  \oplus \frac{1}{\mu_{l}^{b} - \lambda_{b}^{b}(1)}
  \oplus \frac{1}{\mu_{l}^{b} - \lambda_{b}^{b}(2)}
  \oplus \frac{1}{\mu_{l}^{b} - \lambda_{b}^{b}(3)} \right) \nonumber \\
  + \sum_{l=1}^{N_{l}} \left( \mu_{l}^{f} + \mu_{l}^{b} \right), \nonumber
\end{eqnarray}
where $N_{l}$ denotes the total number of cycles, 
$\oplus$ is an exclusive addition operator: one of 
$\lambda_{l}^{f}(1)$, $\lambda_{l}^{f}(2)$, $\lambda_{l}^{f}(3)$, 
$\lambda_{l}^{b}(1)$, $\lambda_{l}^{b}(2)$, $\lambda_{l}^{b}(3)$, 
is chosen for the $l$-th cycle on $path(i,j)$ with the corresponding 
$\mu_{l}^{f}$ or $\mu_{l}^{b}$, 
and $path(i,j)$ denotes a set of cycles with respect to the 
edges of forward or backward direction on the routing path 
between nodes $i$ and $j$. 
There are trade-off relations between the 1st \& 2nd rows of 
the end-to-end delay and the 3rd row of the service load 
in Eq. (\ref{eq_cost}) for minimizing the delivery cost $C$. 
Eq. (\ref{eq_cost}) is rewritten to the sum of cycle-based components 
\[
  \sum_{l=1}^{N_{l}} \left( \sum_{k=1}^{3} \left(
  \frac{w_{l}^{f}(k)}{\mu_{l}^{f} - \lambda_{l}^{f}(k)} 
  + \frac{w_{l}^{b}(k)}{\mu_{l}^{b} - \lambda_{l}^{b}(k)} \right)
  + \mu_{l}^{f} + \mu_{l}^{b} \right), 
\]
where $w_{l}^{f}(1)$, $w_{l}^{f}(2)$, $w_{l}^{f}(3)$, 
$w_{l}^{b}(1)$, $w_{l}^{b}(2)$, $w_{l}^{b}(3)$, 
are the weights obtained from the deformation of Eq. (\ref{eq_cost}).

We can solve the optimal problem for minimizing 
the object function of Eq.(\ref{eq_cost}) 
by using the Newton-Raphson method, 
since the finding at 
$\partial C / \partial \mu_{l}^{f} = 0$ 
(or $\partial C / \partial \mu_{l}^{b} = 0$) 
results in a one-dimensional search for each variable 
$\mu_{l}^{f}$ or $\mu_{l}^{b}$.
We remark that the solution of 
\begin{eqnarray}
  \frac{\partial}{\partial \mu_{l}^{f}}
  \left( \frac{w_{l}^{f}(k)}{\mu_{l}^{f} - \lambda_{l}^{f}(k)} + \mu_{l}^{f}
  \right) \nonumber \\ 
  = \frac{-w_{l}^{f}(k)}{(\mu_{l}^{f} - \lambda_{l}^{f}(k))^{2}} + 1 = 0, 
\nonumber 
\end{eqnarray}
is given by $\mu_{l}^{f} = \sqrt{w_{l}^{f}(k)} + \lambda_{l}^{f}(k)$.
Thus, for minimizing the cost $C$ of Eq.(\ref{eq_cost}), 
it is useful
that the initial value is set as 
\begin{equation}
  \mu_{l}^{f} = \sqrt{w_{l}^{f}(k')} + \max \{ \lambda_{l}^{f}(1), 
  \lambda_{l}^{f}(2), \lambda_{l}^{f}(3) \}, 
\label{eq_initial_value}
\end{equation}
considering the condition (\ref{eq_condition}). 
$k'$ denotes $\argmax \{ \lambda_{l}^{f}(k) \}$. 
Since a triangle has the minimum number of edges to form 
a polygonal cycle, only three variables $\lambda_{l}^{f}(1)$, 
$\lambda_{l}^{f}(2)$, and $\lambda_{l}^{f}(3)$ affect 
to the optimal solution of $\mu_{l}^{f}$.
If we consider the proposed scheme of message-ferries routing 
to any other shaped cycles with more than three edges, 
such as on squares, on chair or sphinx type polygons \cite{Hayashi10}, 
the tuning of service rates 
will be more constrained than the triplets 
in the maximum of Eq. (\ref{eq_initial_value})
for minimizing the cost $C$ of Eq.(\ref{eq_cost}). 
Thus, the triangle cycle is the best choice.
In the above discussion, 
the case of backward direction is similarly treated by 
replacing the superscript from $^{f}$ to $^{b}$.

\section{To be persistent living system}
\subsection{Adaptive recovery for local accidents}
\begin{figure}[htb]
\centering
  \includegraphics[height=32mm]{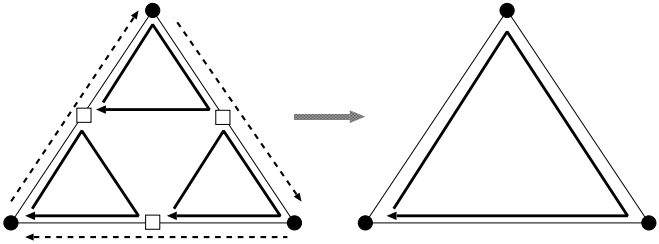}
\caption{Recovery procedure from (left) to (right) 
for a node's accident denoted by white square.
Dashed arrow line shows the move of ferry in changing 
from each cyclic route of bold arrow line to the larger unified one.}
\label{fig_recovery1}
\end{figure}

\begin{figure}[htb]
\centering
  \includegraphics[height=32mm]{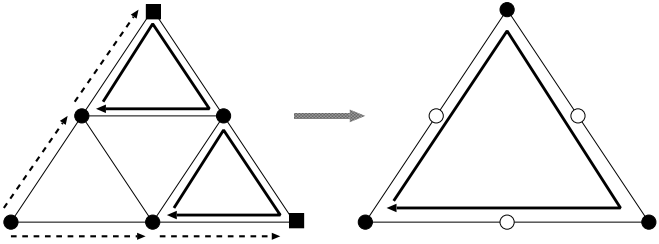}
\caption{Recovery procedure from (left) to (right) 
for a ferry's accident at the bottom left triangle.
Dashed arrow line shows the notification among nodes.
Bold arrow line shows the ferry's cyclic route. 
White circle represents that the node becomes inactive.
}
\label{fig_recovery2}
\end{figure}

We have two strategies of ferry initiation and node initiation 
with monitoring and controlling of the system state 
for a recovery in local accidents.
These situations are illustrated 
in Figs. \ref{fig_recovery1} and \ref{fig_recovery2}.
In the following, although we explain the procedures for the 
clockwise cycles in the 
case of Fig.\ref{fig_msq_cycle}, the same ways can be applied 
for the counterclockwise cycles and for 
the case of Fig. \ref{fig_another_idea}.
When a white square node falls into malfunction and becomes 
non-interactive with a ferry, 
each of the related three ferries detects it in some turnarrounds, 
and moves to the next neighbor black circle node beyond the 
removed white square node, as shown in Fig. \ref{fig_recovery1}.
Then, their ferries begin to move on the larger cycle.

\begin{figure}[htb]
\centering
  \includegraphics[height=32mm]{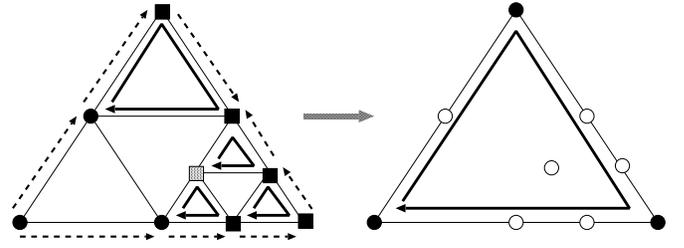}
\caption{Recovery procedure from (left) to (right) 
with hierarchical propagation 
for a ferry's accident at the bottom left triangle.
Dashed arrow line shows the notification among nodes. 
Bold arrow line shows the ferry's cyclic route. 
White circle represents that the node becomes inactive.
}
\label{fig_recovery3}
\end{figure}

When a ferry encounters an accident and stop the moving, 
the black circle nodes that detect it from non-visiting of ferry 
in a time-interval notify the next neighbor black square nodes
to initiate the recovery process, 
as shown in Fig. \ref{fig_recovery2}.
Each of the notified black square nodes negotiates with the 
moving ferries on the cycles of small triangles, 
then a new cycle of the larger triangle 
is created by changing the cyclic routes for the ferries.
In deeper layers, 
the recover process is similarly performed 
as shown in Fig. \ref{fig_recovery3}. 
The gray square node in Fig. \ref{fig_recovery3} 
detects non-visiting of ferry after 
changing the cyclic route, and becomes inactive.
Thus, 
if an accident occurs at a shallow layer, 
the procedures are hierarchically propagated to deeper layers.

\subsection{Update of cyclic paths in the growth}
\begin{figure}[htb]
\centering
  \includegraphics[height=32mm]{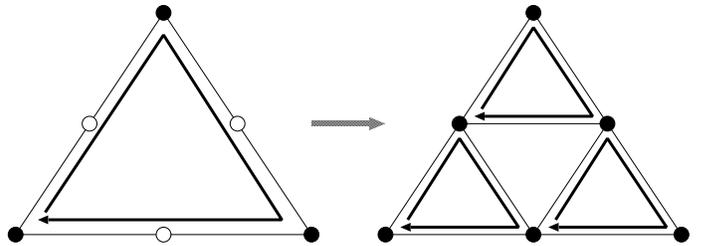}
\caption{Update procedure from (left) to (right) 
by subdivision of triangle face in the network growth.
Bold arrow line shows the ferry's cyclic route. 
From white to black circle, the added node becomes active.
}
\label{fig_update_growth}
\end{figure}

On the other hand, 
we consider the following procedures, 
when new nodes are added by subdivision of triangle face in 
order to be a persistent living system for the growing MSQ network 
and the message-ferries routing on it.
After detecting the addition of nodes by ferries, 
these ferries negotiate with the new nodes to move on 
one of the smaller cycles, as shown in Fig. \ref{fig_update_growth}.
Such procedures can be performed at any layers of triangle faces 
and even simultaneously in some distant parts.

\begin{figure}[htb]
\centering
  \includegraphics[height=73mm]{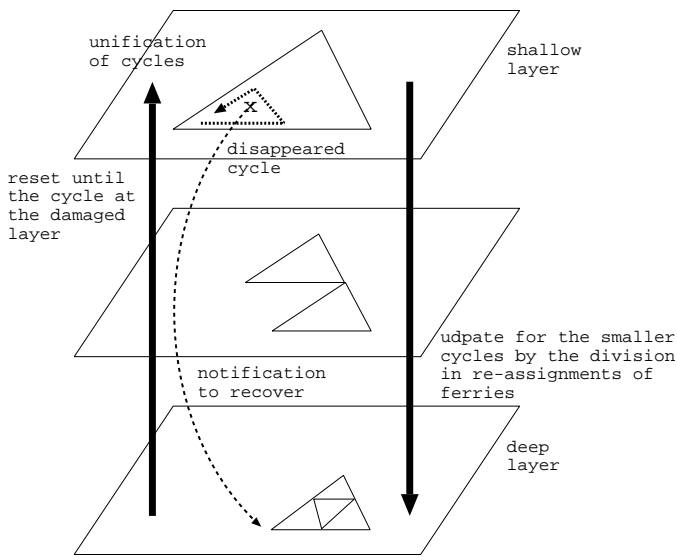}
\caption{Unification and re-division of triangle cycles.
}
\label{fig_hierarchical_recovery}
\end{figure}

The update process is used to the hierarchical recovery for a ferry's 
accident, when nodes work correctively.
The white inactive nodes on the 2nd layer 
at the right of Fig. \ref{fig_recovery3} act same as adding new nodes 
at the left of Fig. \ref{fig_update_growth}, then the related 
message-ferries negotiate with these nodes to move on smaller cycles
at the right of Fig. \ref{fig_update_growth}.
After the change of routes, the inactive nodes on the 3rd layer begin 
to return the state in a right-down triangles at the right of 
Fig. \ref{fig_recovery3}.
These procedures propagate to deeper layers hierarchically 
with the unification and division of ferry's routes 
from transiently resetting and re-creating the cycles on 
the layers back to the damaged layer by the accident, 
as shown in Fig. \ref{fig_hierarchical_recovery}.
The resetting and re-creating seem to be wasteful, however 
using such consistent simple ways 
is better especially for unpredictable situation 
in maintaining the minimum performance.
Because extremely complex procedures and controls can be avoided.

\section{Conclusion}
We have proposed a design principle in CAS 
for a DTN routing based on 
autonomously moving message-ferries 
on a special structure of MSQ network 
self-organized by iterative subdivision of faces.
In the subdivision, 
nodes are located as balancing the amount of communication requests
as possible 
which requests are generated or received at a node 
according to population in its territory defined by the nearest access.
By considering 
the correspondence between each cycle on a triangle 
and a moving of message-ferry, we realize a reliable and efficient 
message-ferries routing.
In the routing, a relay of cyclic message-ferries, the bounded path 
length by the $t$-spanner property, and the adaptive face routing algorithm 
are key factors.
As stochastic processes, 
we have considered the arrival of communication request 
and the service of transfer 
in the relay of cyclic message-ferries. 
The tandem queue model in a queueing theory is applied for this 
problem setting.
We have derived a calculation method of the optimal solution 
for minimizing the delivery cost defined by a sum of 
the end-to-end delay and the service load 
in a trade-off relation. 
Moreover, we have considered recovery procedures for local 
accidents, 
and update procedures of ferry's cyclic routes
in an incremental growth of MSQ network. 
These collective adaptive configuration and procedures 
will be evaluated 
in numerical simulations for more detailed design.

We mention some further studies. 
Our proposed method can be applied to a transport logistic system 
(e.g. including by vehicles or small flying devices) 
in both normal and emergent situations on a wide area. 
However, 
how to define a processing unit of communication or transportation 
requests is an issue.
In addition, 
temporal disconnecting cases at simultaneous and many parts 
may be discussed by extending our 
approach in considering disaster situations.
These challenges will be useful for developing 
a resilient network as CAS in theoretical and practical 
points of views \cite{Hollnagel06,Zoli12}.


\section*{Acknowledgment}
This research is supported in part by
Grant-in-Aide for Scientific Research in Japan, No.21500072.



%

\end{document}